\documentstyle[12pt]{article}

\renewcommand{\a}{\alpha}
\renewcommand{\b}{\beta}

\renewcommand{\d}{\delta}

\newcommand{\ep}{\epsilon}
\newcommand{\G}{\Gamma}
\newcommand{\k}{\kappa}
\newcommand{\la}{\lambda}
\newcommand{\La}{\Lambda}
\newcommand{\om}{\omega}

\newcommand{\si}{\sigma}
\newcommand{\Si}{\Sigma}
\newcommand{\th}{\theta}
\def\Th{\Theta}

\def\sib{{\bar \sigma}}
\def\thb{{\bar \theta}}

\newcommand{\p}{\partial}

\newcommand{\non}{\nonumber\\}

\newcommand{\beq}{\begin{equation}}
\newcommand{\eeq}{\end{equation}}
\newcommand{\beqa}{\begin{eqnarray}}
\newcommand{\eeqa}{\end{eqnarray}}

\newcommand{\refeq}[1]{(\ref{#1})}

\newcommand{\plb}[1]{{ \it Phys.~Lett.}~{\bf {#1B}}}

\newcommand{\npb}[1]{{ \it Nucl.~Phys.}~{\bf B{#1}}}

\newcommand{\cmp}[1]{{\it Comm.~Math.~Phys}~{\bf {#1}}}

\def\half{{\mbox{\small  $\frac{1}{2}$}}}

\def\det{{\rm det}}

\def\Mb{{\bar M}}

\def\thb{{\bar \theta}}

\def\sib{{\bar \si}}

\def\vy{{\vec y}}
\def\vY{{\vec Y}}
\def\vtau{{\vec \tau\,}}
\def\vom{{\vec \om}}

\def\ad{{\dot \a}}
\def\bd{{\dot \b}}

\def\sui{SU$_I$(2) }
\def\fp{F_{+\mu\nu}}
\def\tpp{t_+^{\mu\nu}}
\def\som{\om^{\mu\nu}}

\begin{document}

\baselineskip 20pt plus 2pt
\begin{titlepage}
\renewcommand{\thefootnote}{\fnsymbol{footnote}}
\begin{flushright}
\parbox{1.5in}
{
 UW-IFT-27/96\\
 December, 1996\\
hep-th/9706134\\}
\end{flushright}
\vspace*{.5in}
\begin{centering}
{\Large \bf From Seiberg-Witten invariants to topological Green-Schwarz
        string \footnote{ 
Work supported in part
by Polish State Committee for Scientific Research (KBN) and EC under the
contract ERBCIP-DCT94-0034.}}\\
\vspace{2cm} 
{\large        Jacek Pawe\l czyk}\\
\vspace{.5cm}
        {\sl Institute of Theoretical Physics, Warsaw University,\\
        Ho\.{z}a 69, PL-00-681 Warsaw, Poland.}\\
\vspace{.5in}
\end{centering}
\begin{abstract}
In this note we describe the physics of equivalence of the Seiberg-Witten
invariants of 4-manifolds and certain Gromov-Witten invariants defined by
pseudo-holomorphic curves. We show that physics of the pseudo-holomorphic
curves should be governed by the N=2 Green-Schwarz string. 
\end{abstract}
\vfill
\end{titlepage}

\renewcommand{\thefootnote}{\arabic{footnote}}

\section{Introduction} 
In the recent paper \cite{taubes} Taubes announced the equivalence of the
Seiberg-Witten  
invariants \cite{fourm} of symplectic 4-manifolds \cite{donald,dusa2} and
certain Gromov-Witten 
invariants defined by pseudo-holomorphic curves \cite{gromov,topsigma,dusa}.
Besides its main application to topology the result has also interesting 
physical aspects.
It sounds like the long standing expectations that in some cases gauge
field configurations should be well approximated by strings. In our case the
string theory would be the topological theory of pseudo-holomorphic curves.
Of course the problem lies far apart from the 1/$N_c$ expansion
\cite{thooft,gross} because there is nothing like $N_c$ here. 
It shall appear that the localization is strictly related to properties of
topological field theory. Anyway the mechanism of localization has 
an interesting
physical background and moreover it shows relations to the N=2 Green-Schwarz
string \cite{gsw}. These are the reasons why we have decided to base  our
discussion on physical N=2 SUSY theories which by 
``twisting''\footnote{I would like to thank R.Stora for seeding an uneasiness
into my naive believe in ``twisting''.} are related to topological field
theories \cite{witten:top}.  

We start with the N=2 SUSY SU(2) Yang-Mills theory in $R^4$ and show that
after inclusion of Feyet-Iliopoulos (FI) term it localizes on
pseudo-holomorphic curves. In the next section we show the relation to the N=2
Green-Schwarz string.

\section{From Seiberg-Witten theory to
pseudo-holomo-rphic curves}

We start from the N=2 SUSY SU(2) Yang-Mills theory which is a basis for the 
Seiberg-Witten  theory of invariants of 4-manifolds \cite{fourm}.
We recall that the moduli space of the N=2 SUSY SU(2) Yang-Mills theory
\cite{sw}  
contains two singular points. At these points the low energy effective 
theory is N=2 
SUSY U(1) theory coupled to an additional massless matter
(monopoles or dyons) in the form of the  N=2 hypermultiplet.  
Hence the low energy fields are: the vector multiplet (gauge field
$A_\mu$, \sui doublet of fermions $\la^i_\pm$, scalar $\phi$ plus \sui triplet
of auxiliary fields $\vY$)\footnote{Our conventions are:
\sui spinors doublets respect the so-called symplectic Majorana condition
$\psi^i_\pm=\ep^{ij}(i\si^2)(\psi^j_\pm)^*,\quad i,j=1,2$, where $\psi^i_\pm$
are spinors of SU$_\pm$(2) and the Lorenz group is $SO(4)=SU_+(2)\times
SU_-(2)/Z_2$. Also $\si^\mu=(1,i{\vtau})\quad \sib^\mu=(1,-i{\vtau})$ and 
$\si^{\mu\nu}=\frac{i}{4}(\si^\mu\sib^\nu-\si^\nu\sib^\mu)$, $\sib^{\mu\nu}=
\frac{i}{4}(\sib^\mu\si^\nu-\sib^\nu\si^\mu)$.}
and the 
hypermultiplet (\sui doublet of scalars $M^i$, two fermions $\psi_\pm$).
We shall write down only the relevant terms of the low energy Lagrangian.
Thus we drop  fermions and the boson $\phi$ which has trivial vacuum.
\beq
L=\frac{1}{4}F_{\mu\nu}^2-\frac{1}{2}(\vY)^2+
D_\mu {\bar M}^iD^\mu M^i-\vY\,\Mb^i\vtau^{ij}M^j
\label{lag}
\eeq 
Because of the future application to topological field theory we have written
the Lagrangian 
\refeq{lag} in the Euclidean space.

It is  known that N=2 SUSY admits also Feyet-Iliopoulos (FI) term
of the form 
\beq
-\vY \vom
\label{fi}
\eeq 
The term goes with \sui triplet of coupling 
constants $\vom$ of the dimension of mass squared. 
The auxiliaries are members of the vector multiplet so FI term does
not break U(1) explicitly but can break it spontaneously. It can also
spontaneously break SUSY.  

Let us examine the classical vacuum of the model. After solution of
the equation of motions for auxiliaries from \refeq{lag} plus \refeq{fi} 
we get that the potential is 
\beq
V= \half (\Mb^i\vtau^{ij}M^j-\vom )^2.
\eeq 
thus the vacuum respects $\vY=\vom-\Mb^i\vtau^{ij}M^j=0$. This breaks the U(1)
completely. In this case standard arguments lead to the conclusion
that there  exist strings as topological solitons. The width of the string is
of the order $1/|\om|$. In the limit $|\om|\to\infty$ all classical field
configurations are localized along these strings.\footnote{ 
Of course in the  case of Euclidean theory instead of static strings 
we shall have 2d surfaces as topological solitons. In the course of 
the paper we have decided  to keep the physical terminology hoping that it
will not lead to confusions. 
Moreover we shall deal  only with SUSY transformations generated by $\xi_+$ 
because we want to twist $SU_+(2)$ subgroup of the (Euclidean Lorentz) 
group $O(4)$. } 

The supersymmetry transformations in the nontrivial (bosonic)
background are as follows: 
\beqa
\d\la^i_+&=&i( \vtau^{ij}\vY+F_{ +\mu\nu}\sib^{\mu\nu}\d^{ij})\xi^j_+
\label{lap}\\ 
\d\la^i_-&=&i( \vtau^{ij}\vY+F_{ -\mu\nu}\sib^{\mu\nu}\d^{ij})\xi^j_-\\ 
\d \psi_+&=&-i\sib^\mu D_\mu M^i\ep^{ij}\xi^j_-
\\
\d \psi_-&=&-i\si^\mu D_\mu M^i\ep^{ij}\xi^j_+
\label{psim}
\eeqa
where $F_\pm=\half (F\pm * F)$. Vanishing of (\ref{lap},\ref{psim}) for some
parameters $\xi^i$ yields Bogomolny type conditions. 
This kind of conditions has
been recently thoroughly studied 
within the context of the so-called extremal p-branes for string inspired
Lagrangians \cite{duff}. 
In the context of topological  field theory 
the standard choice for a constant spinor  is 
$\xi^{i\ad}_+=\ep^{i\ad}\eta$ with real $\eta$ ($\ad$
is the index of spinor representation of SU$_+$(2)). 
Then the equation of motions following from vanishing of
(\ref{lap},\ref{psim}) read 
\beqa
0&= &\fp+Y_{\mu\nu},\quad \mu,\nu=0,1,2,3\non
0&=& \si^\mu_{\a\bd}D_\mu M^\bd 
\label{sw-eq}
\eeqa
where $Y_{\mu\nu}=\om_{\mu\nu}-\Mb\sib_{\mu\nu}M$ is self dual tensor defined
by $Y^{0i}=Y^i/2$. 
These are famous Seiberg-Witten equations perturbed by the Feyet-Iliopoulos
term $\vom$. 
This perturbation  has been 
 considered previously by many authors \cite{fourm,taubes2,marcolli}.

Moreover the configurations respecting \refeq{sw-eq} break only half (two out
of four) of the
supersymmetries $\xi^i_+$. One is $\xi^{i\ad}_+=\ep^{i\ad}\eta$. 
In order to see the another  one let us denote
the  operator standing on the r.h.s. of \refeq{lap} by $\Th=(\fp
\sib^{\mu\nu}\otimes 1+ \vY\, 1\otimes\vtau )$. The operator is hermitian and
traceless. It also anticommutes with the symplectic Majorana operator
$i\tau^2\otimes i\tau^2$ (see the last footnote). It follows that 
zero eigenvalues go in pairs. Thus
there must be two of them and we can choose them to be the eigenvalues of
$i\tau^2\otimes i\tau^2$.  One of them is just $\xi^{i\ad}_+=\ep^{i\ad}\eta$.
The second SUSY
transformation \refeq{psim} does not give any new condition. 
To show it we apply the chiral Dirac operator 
$\sib^\mu D_\mu$ on the r.h.s. of \refeq{psim}.
\beq
\sib^\mu D_\mu\d \psi_-=[D^2-F_{\mu\nu}\sib^{\mu\nu}]M^i\ep^{ij}\xi^j_+ 
\eeq
By the formula \refeq{lap}, vanishing of $\d \la^i_+$ transforms this to 
\beq
[D^2M^i-\vY\vtau^{ik}M^k]\ep^{ij}\xi^j_+ 
\eeq
The expression in square brackets is just the equation of motion for the
scalar $M$ following from \refeq{sw-eq} or \refeq{lag}. 
Thus $\sib^\mu D_\mu \d\psi_-=0$. 
We expect that in generic case this implies $\d\psi_-=0$.
We conclude that out of four components of SUSY generators two of them are
preserved by the background which respect \refeq{sw-eq}.
 
The two broken SUSY generators have nice physical interpretation \cite{polch}.
They are Goldstone fermions of partially broken supersymmetry  and they
become  
fermionic coordinates of the soliton. As we shall see the effective
description of the solitonic string in the zero width limit  will 
be given by the Green-Schwarz action. This string has
 two physical space-time spinors. As we shall see these spinors are 
exactly the two SUSY generators that have been broken. 
This will be one of the argument for the
identification of the Green-Schwarz string as the string describing solutions
of \refeq{sw-eq} in the limit $|\om|\to \infty$.

Here we shall recall some properties of the physical stringy solitons in 
order to build-up an intuition what happens in the Euclidean case.
We notice that the solitons are 
BPS states thus if they are parallel  they do not interact. 
This means that multiple
string configurations are also solutions of the equation of motions. For the
detailed discussion of this point see \cite{taubes:vort}. 
The width of a single string is of the order $1/\sqrt{|\om|}$ and shrinks to
zero 
when  $|\om|\to \infty$. 
This limit in quantum field theory \refeq{lag} should be
somehow ill 
defined as we deal with UV region of non asymptotically free field theory. 
This is probably the 
source of troubles of the Green-Schwarz string. 
The troubles vanish when one goes to topological field theory. 
Thus we expect no troubles with the $|\om|\to \infty$ limit. In this limit 
the topological configurations solving \refeq{sw-eq} should be localized 
on a 2d surfaces and moreover whenever the notion of being ``parallel'' 
can be 
extended to curved 4-manifolds we should expect multi-soliton solutions.
The last situation can occur for the surfaces being tori.

Now we shall derive the equations determining  the surface
on which solutions to \refeq{sw-eq} localizes when $|\om|\to \infty$. 
The position of the string is determined by the equation
$M^i(x)=0$. We dwell for a moment on the physical, static configurations. 
Thus we pick one of the coordinate (say $x^0$) to be the time and 
choose $A_0=0$ gauge.
Then Eqs.\refeq{sw-eq} have the form
\beq
{\vec B}=-\frac{1}{2}(\vom+\Mb\vtau M)
\eeq
With the appropriate boundary conditions at infinity 
the direction of the string 
(i.e. the line $X^m(\si)$, which respects 
$M^i({\vec x})|_{\vec x}={\vec X}(\si)=0$) 
is given by $\vom$ i.e. the string equation is 
$\p_\si{\vec X}=\vom$ where $\si$ parameterize the string. 
Fields $M^i,\;{\vec A}$ have  non-trivial
behavior on the  surface perpendicular  to $\vom$.
Now we go the  general case. 
Let the surface of zero locus of 
$M^i$ be given by an immersion $X:\Si\to M=R^4$. 
Thus if we pull back first of the 
Eqs.\ref{sw-eq} on $X$ we get
\beq
F^+_{\mu\nu}\tpp=\som\tpp=const\neq 0\mbox{  on }\Si
\label{surf1}
\eeq
where
$\tpp=t^{\mu\nu}+*t^{\mu\nu}$ and $t^{\mu\nu}\equiv \ep^{ab}\p_a X^\mu \p_b
X^\nu/\sqrt{g}$ ($a,b$ indexes coordinates on
$\Si$, $g=\det (g_{ab}),\; g_{ab}=\p_a X^\nu\p_b X^\nu$ 
is the induced metric). The form $\om$ determines an almost complex 
structure $J_\om$ on our space-time $M$ as well as on the pull-back 
tangent bundle
$X_*TM$. Using  $J_\om$ one can decompose the space of self-dual forms 
as follows 
\beq
X_*\La^2_+M=X_*\La^{(1,1)}\oplus X_*\La^{(2,0)}\oplus X_*\La^{(0,2)}.
\eeq
Moreover we  choose a hermitian metric on $M$. 
In this metric all components of the above sum are orthogonal to each other.
The equation \refeq{surf1} implies 
\beq
\tpp=\frac{\som}{|\om|}+...
\label{surf}
\eeq
where the dots denotes the undetermined 
contribution of the $X_*\La^{(2,0)}$ and $ X_*\La^{(0,2)}$ forms.
Now we try to match with the just discussed static case. 
We pick up the gauge $X^0\propto t$. In this gauge ${\vec X}$ are functions
of $\si$ only. Now comparing $\p_\si{\vec X}=\vom$ with \refeq{surf}
one sees that the dots in the latter do  not contribute.
This fixes \refeq{surf} (without dots) 
as the equation for the surface on which the 
solutions of the deformed Seiberg-Witten equations localizes in the limit
$|\om|\to \infty$. 
The equations
\refeq{surf} are, in fact, equations for pseudo-holomorphic curves. We can see
it multiplying both sides of \refeq{surf} by $\p_a X^\nu$. 
Using the definition
of the induced metric $g_{ab}=\p_a X^\nu\p_b X^\nu$ we  get 
\beq
\frac{\som}{|\om|}\p_a X^\nu=\frac{\ep_a^{\; b}}{\sqrt{g}}\p_b X^\mu
\eeq
Pseudo-holomorphic curves defines the so-called Gromov-Witten invariants. 

These simple  arguments do not give the precise relation between 
the Seiberg-Witten theory and the Gromov-Witten theory 
 but show the physics of the localization. For the mathematical
description we refer the reader to the original papers of 
Taubes \cite{taubes}. Nevertheless one obtains a simple 
understanding of certain
properties of the relation e.g. the necessity for multiple-covered tori or 
the use of disconnected surfaces. The
physical approach presented above allows also to  derive the string theory 
relevant for this problem. We shall dwell on this problem in
the next section.
 
\section{Green-Schwarz string}

As we have showed in the previous section, the  Seiberg-Witten theory
\refeq{sw-eq}, in the limit $|\om|\to\infty$,
localizes on the pseudo-holomorphic 
curves \refeq{surf}.
Here we would like to concentrate on the physical string
 theory which is directly related to the previously studied field theory.  We
shall show that the string theory is familiar N=2 Green-Schwarz string
\cite{gsw}. Of 
course the model is quantum mechanically 
ill-defined for 4d space-times what is presumably reflection of the discussed
UV sickness of the N=2 SUSY U(1) field theory. As we
are interested in topological field theory this limit 
should not produce any problems. 

The are several arguments leading to the  N=2 Green-Schwarz string.
First of all the string has  massless spectrum exactly the same as 
that of the N=2 SUSY
U(1) coupled to one N=2 hypermultiplet. Moreover as we shall  
 show below  the spinors of the N=2 Green-Schwarz string are
the  Goldstone modes of the partially broken supersymmetry of the 
field theory. Together with the superysmmetry of the theory this 
uniquely identifies the N=2 Green-Schwarz string as 
the effective theory describing the solitonic string of the previous section.

To match with field theory description of the previous section we must
slightly modify the original formulation of the Green-Schwarz string. The
modification revels the \sui structure of the theory. One should expect this
because the equation of string world-sheet surface \refeq{surf} depends
explicitly on $\vom$. 
The N=2 Green-Schwarz string consists of the Nambu-Goto term and the
Wess-Zumino term $L_{WZ}$:
\beq
S=\int d^2z\{\half\;\sqrt{g}+L_{WZ}\}
\label{string}
\eeq
where $g_{ab}=\Pi_a^\mu\Pi_b^\mu$,  $\Pi_a^\mu=\p_a
X^\mu-i(\thb^i_+\sib^\mu\p_a\th_-^i + 
\thb^i_-\si^\mu\p_a\th_+^i)$ ($i=1,2$). The famous Wess-Zumino
term 
is responsible for the existence of the local $\k$-symmetry.
The two terms transform differently under \sui: 
the kinetic term is a singlet, 
the Wess-Zumino term is a component of a triplet.
The easiest way to produce the Wess-Zumino term is to construct closed
3-forms in the 
target superspace. It appears that they form a \sui triplet
\def\vOm{{\vec \Omega}}
\beq
\vOm=\Pi^\mu\,d\thb_+^i\sib^\mu d\th_-^j\vtau^{ij}
\eeq
The forms $\vOm$ are  exact thus we define triplet of the Wess-Zumino 2-forms
$\vOm_{WZ}$ by $\vOm=d\vOm_{WZ}$. 
The proposed Wess-Zumino term is:
\beqa
L_{WZ}&=&\vOm_{WZ}\vy\\ 
&=&-\half\ep^{ab}[\p_a X^\mu-\frac{i}{2}(\thb^i_+\sib^\mu\p_a\th_-^i +
\thb^i_-\si^\mu\p_a\th_+^i)]\; (\thb^k_+\sib^\mu\p_b\th_-^l -
\p_b\thb^k_+\sib^\mu\th_-^l)\,\vtau^{kl}\,\vy\nonumber
\label{wz}
\eeqa
where $|\vy|=1$. The vector $\vy$ will be determined by comparison 
with the field theory of the previous section.
The standard Green-Schwarz string action is recovered for
$\vy_{gs}=(0,0,1)$. 
The global SUSY transformation is not changed by this modification,
\beqa
\d X^\mu&=&-i(\thb_+^i\sib^\mu\xi^i_- + \thb_-^i\si^\mu\xi^i_+)\\
\d\th^i_+&=&\xi^i_+\\
\d\th^i_-&=&\xi^i_-
\eeqa
but the $\k$-symmetry is modified:
\beqa
\d X^\mu&=&i(\thb_+^i\sib^\mu\d\th^i_- + \thb_-^i\si^\mu\d\th^i_+)\\
\d\th^i_+&=&i(\vy\,\vtau^{ij}+ \d^{ij}\G_+)\kappa_+^j\\
\label{ksymm}
\eeqa
and similarly for $\d\th^i_-$. 
In the above
\beq
\G_+=\frac{\ep^{cd}\Pi_c^\mu\Pi^\nu_d}{\sqrt{g}}\sib^{\mu\nu}
\label{gamma}
\eeq
We note here that $\sib^{\mu\nu}$ is self-dual so it picks up only the
self-dual part of  the tensor on the r.h.s. of \refeq{gamma}. For
$\th_\pm^i=0$ we have $\G_+=\half\tpp\sib^{\mu\nu}$. 
This will be crucial for comparison with field theory.

Now we connect this picture with the previous field theory considerations.  We
recall that the center of the solitonic string is defined by $M^i=0$, so at
this point the first Seiberg-Witten equation gives $\fp=\om_{\mu\nu}$ thus, by
\refeq{surf}, 
also  
$\fp/|\om|= \half t_{+\mu\nu}$. Inserting this back to \refeq{ksymm} 
and identifying $|\om|\vy=\vY|_{M^ib=0}= \vom$  we
realize 
that  the latter coincides with field theory transformation \refeq{lap} taken
at the 
string world-sheet. This means that  the  spinor which can be gauged away by
\refeq{ksymm} is exactly the field theory
spinor of preserved SUSY transformation \refeq{lap}. 
This is in perfect agreement with physical intuition that only broken
generators of a symmetry constitute the physical degrees of the string
\cite{polch}.

Thus the theory \refeq{string} with the Wess-Zumino term given by \refeq{wz}
and $\vy=\vom/|vom|$ 
is the seeking effective string theory describing 
N=2 SUSY U(1) fields theory
string.  

There are several things which should be farther clarified and worked out.
Among them there is the topological field theory behind the above
N=2 Green-Schwarz string and its comparison with the theory of
pseudo-holomorphic curves \refeq{surf}.
 We shall devote our future publication to these subjects.

\end{document}